%% file: main.tex
\documentclass[10pt,twocolumn,letterpaper]{article}

\usepackage{cvpr}
\usepackage{algorithm,algorithmic}
\usepackage{diagbox} 
\usepackage{hhline}
\usepackage{caption}
\usepackage{bbding}
\usepackage{booktabs}
\usepackage{textcomp}
\usepackage{threeparttable}
\usepackage{graphicx}
\usepackage{multirow,amsmath,epsfig,amsfonts,amssymb,psfig,graphics,psfrag,theorem,calc,url,bm}
\usepackage{makecell}
\usepackage{booktabs}
\usepackage{tikz}
\usetikzlibrary{quantikz}
\usepackage{mathtools}

\def\blue{\color{blue}}
\def\red{\color{red}}
\usepackage{colortbl}
\definecolor{orange}{RGB}{255,107,0}
\def\blue{\color{blue}}

\def\red{\color{red}}

\newcommand\bA{\ensuremath{{\bm A}}}
\newcommand\bB{\ensuremath{{\bm B}}}

\newcommand\bD{\ensuremath{{\bm D}}}
\newcommand\bE{\ensuremath{{\bm E}}}

\newcommand\bG{\ensuremath{{\bm G}}}

\newcommand\bP{\ensuremath{{\bm P}}}

\newcommand\bR{\ensuremath{{\bm R}}}
\newcommand\bS{\ensuremath{{\bm S}}}
\newcommand\bT{\ensuremath{{\bm T}}}
\newcommand\bU{\ensuremath{{\bm U}}}

\newcommand\bX{\ensuremath{{\bm X}}}

\definecolor{orange}{RGB}{255,107,0}
\def\blue{\color{blue}}
\def\red{\color{red}}

\input{preamble}
\definecolor{cvprblue}{rgb}{0.21,0.49,0.74}
\usepackage[pagebackref,breaklinks,colorlinks,allcolors=cvprblue]{hyperref}

\def\paperID{*****} 
\def\confName{CVPR}
\def\confYear{2026}

\title{Spectral Super-Resolution via Adversarial Unfolding \\
and Data-Driven Spectrum Regularization: \\
From Multispectral Satellite Data to NASA Hyperspectral Image}

\author{Si-Sheng Young\\
Institute of Computer and Communication Engineering\\ National Cheng Kung University\\
Tainan, Taiwan (R.O.C.)\\
{\tt\small  q38121509@gs.ncku.edu.tw}
\and
Chia-Hsiang Lin\\
Department of Electrical Engineering\\
National Cheng Kung University\\
Tainan, Taiwan (R.O.C.)\\
{\tt\small chiahsiang.steven.lin@gmail.com}
}

\begin{document}
\maketitle
\input{sec/0_Abstract}    
\input{sec/1_Intro}
\input{sec/2_Method}

\input{sec/3_Experiment}

\input{sec/4_Conclusion}

\input{sec/X_Supple}
{
    \small
    \bibliographystyle{ieeetr}
    \bibliography{main}
}

\end{document}

%% file: sec/0_Abstract.tex
\begin{abstract}
The European Space Agency's Sentinel-2 satellite provides global multispectral coverage for remote sensing (RS) applications.
However, limited spectral resolution (12 bands) and non-unified spatial resolution (60/20/10 m) restrict their practicality.  
In contrast, the high spectral-spatial resolution sensor (e.g., NASA's AVIRIS-NG) covers only the American region due to practical considerations.
This raises a fundamental question: ``Can a global hyperspectral coverage be achieved by reconstructing Sentinel-2 data to NASA hyperspectral images?''
This study aims to achieve spectral super-resolution from 12-to-186 and unify the spatial resolution of Sentinel-2 data to 5 m.
To enable a reliable and efficient reconstruction, we formulate a novel deep unfolding framework regularized by a data-driven spectrum prior from PriorNet, instead of relying on implicit deep priors as conventional deep unfolding does.
Moreover, an adversarial term is integrated into the unfolded architecture, enabling the discriminator to guide the reconstruction in both the training and testing phases; we term this novel concept unfolding adversarial learning (UAL).
Experiments show that our UALNet outperforms the next-best Transformer in PSNR, SSIM, and SAM, while requiring only 15\% MACs and 20 times fewer parameters.  
The associated code will be publicly available at \href{https://sites.google.com/view/chiahsianglin/software}{\textit{https://sites.google.com/view/chiahsianglin/software}}.
\end{abstract}


%% file: sec/1_Intro.tex
\section{Introduction} \label{sec:Intro}
\begin{figure}[t]
    \centerline{\includegraphics[width=0.5\textwidth]{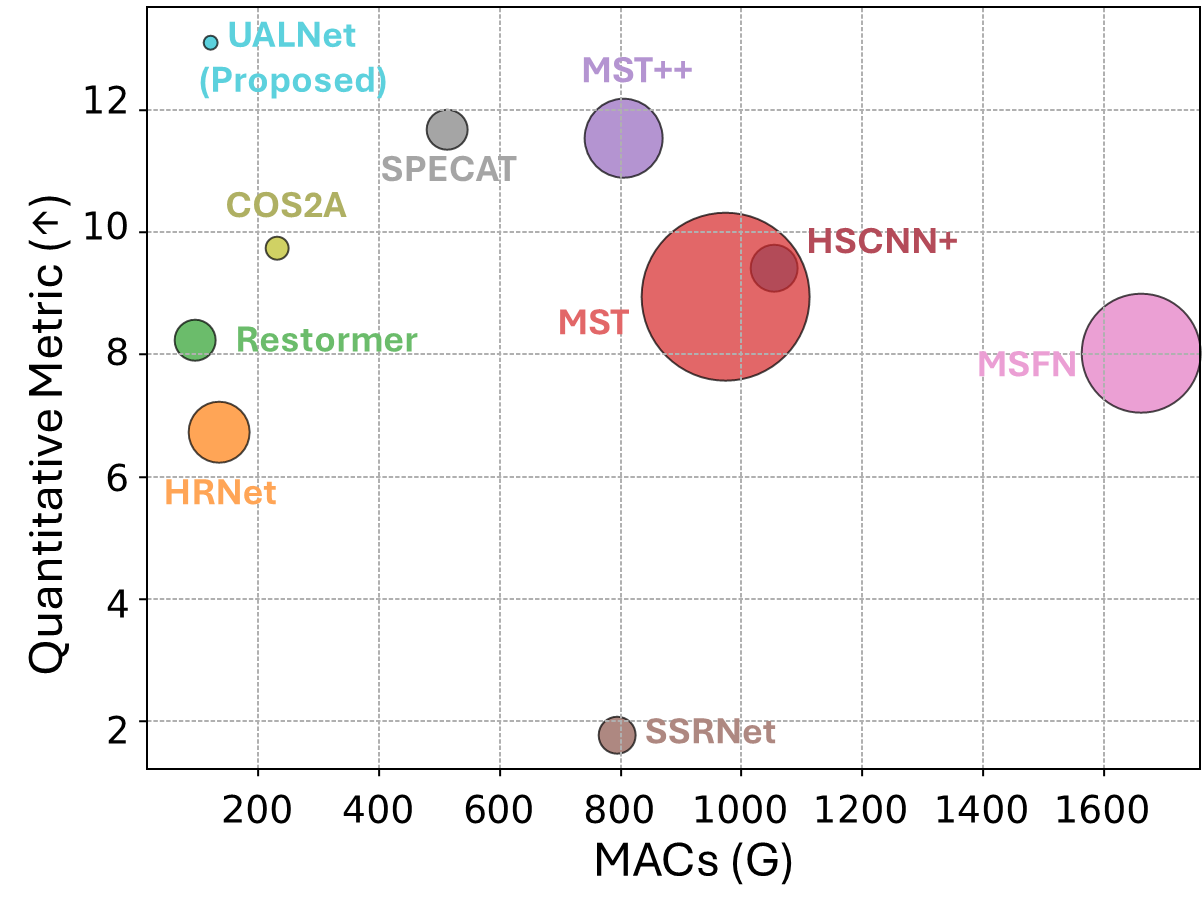}}
    \caption{Performance-Params-MACs comparisons with spectral/spatial reconstruction models. 
    The horizontal axis is computational complexity (measured in MACs ), the vertical axis indicates performance [reported as PSNR-over-SAM ratio $(\uparrow)$ to consider both spatial and spectral fidelities], while the circle radius corresponds to the network parameters (memory cost). 
   When the performance stems from sophisticated architecture and model depth, it results in prohibitive computational complexity and parameters. 
    Conversely, the proposed unfolding adversarial learning network (UALNet) achieves the highest performance with substantially lower MACs and Params with the explainable architecture.
    }\label{fig: flops}
\end{figure}
Supported by the Copernicus programme, the European Space Agency’s Sentinel-2 constellation provides global remote sensing (RS) imagery for Earth observation \cite{S2data}.
Operating the Sentinel-2A and Sentinel-2B satellites, the system enables a fine temporal resolution with a minimum global revisit period of approximately five days.
Furthermore, its broad spectrum range (i.e., approximately from 440 to 2200 nm) facilitates numerous RS applications \cite{SSSS}, including precision agriculture \cite{S2_agriculture}, land and natural disaster monitoring \cite{S2_land}, mangrove mapping \cite{10285517}, and change detection \cite{S2_CD}.
However, due to inherent hardware considerations, such as atmospheric absorption and signal-to-noise ratio (SNR), Sentinel-2's multispectral data comprise merely 12 effective spectral bands with various spatial resolutions, i.e., ground sampling distance (GSD) of 10 m, 20 m, and 60 m, limiting their practicality.
Even when the multiresolution issue is alleviated, the 12-band spectral signatures remain sensitive to deviations and are less reliable for materials identification compared to typical hyperspectral images (HSIs) \cite{fu2019hyperspectral,scheibenreif2023masked}.

In contrast, a typical hyperspectral imaging system delivers HSIs with high spectral resolution and a uniform spatial resolution.
A representative example is NASA’s airborne visible/infrared imaging spectrometer (AVIRIS).
For instance, the AVIRIS-Classical (AVIRIS-C) \cite{AVIRISC} instrument offers calibrated HSIs with spectrum wavelengths from 400 nm to 2500 nm at a GSD of approximately 10 m, yielding 224 contiguous high-resolution bands (around 180 effective bands after removing water vapor absorption bands).
Furthermore, the AVIRIS-Next Generation (AVIRIS-NG) \cite{AVIRISNG} system, covering 380 nm to 2510 nm under a spectral sampling interval of 5 nm and about 5 m GSD, has been developed as a successor of AVIRIS-C. 
Unfortunately, such an airborne hyperspectral system operates primarily over the American region and hence cannot support global coverage, as the Sentinel-2 constellation does.
Given their highly overlapping spectral wavelengths and comparable spatial resolution, a natural question arises: ``\textit{Can a global hyperspectral coverage be achieved by reconstructing Sentinel-2 data to NASA hyperspectral images?}''
Successfully addressing this challenging task is expected to provide significant value and impact for RS applications.
A recent study \cite{COS2A} has demonstrated the feasibility of a challenging 12-to-172 spectral super-resolution (SSR) and spatial resolution unification (SRU) to 10 m GSD  via the integration of convex optimization and deep learning.
Nevertheless, although the spatial resolution has been enhanced to be GSD 10 m, the reconstructed HSIs (i.e., the estimation from Sentinel-2 data) may still suffer from the so-called mixed pixel phenomenon (MPP) \cite{zeng2024unmixing,yu2024unmixing,HyperCSI}, meaning that each hyperspectral signature represents a mixture of multiple distinct materials.
Conversely, MPP is expected to be substantially mitigated when GSD is equal to or below GSD of 5 m \cite{10641882,10462214}, motivating us to pursue a finer spatial scale.
Meanwhile, achieving a higher spectral resolution is desirable to facilitate more precise material identifications.
Thus, this study aims to advance further the state-of-the-art reconstruction, i.e., 12-to-186 SSR and the 5 m GSD SRU.

Notably, the additional 2$\times$ spatial SR (from 10 m to 5 m GSD) in our target reconstruction is far from being as simple as a typical computer vision task. 
In fact, it substantially elevates the level of difficulty.
When performing Sentinel-2-to-AVIRIS reconstruction at 10 m GSD, the high-resolution Sentinel-2 bands provide strong spatial guidance.  
Algorithms can focus on spectral reconstruction preliminarily, and subsequently treat the estimated HSI and high-resolution Sentinel-2 bands as a typical data fusion \cite{zhao2023ddfm,CNMF} problem instead of a single-image spectral-spatial SR \cite{MSFN,DeepHSI}.
On the other hand, one may expect that a sophisticated network or generative model, such as a Transformer-based architecture \cite{MST++}, diffusion model \cite{10947187,zhang2025diffusion,fu2025univg}, autoregressive \cite{xiong2024autoregressive}, or flow-based models \cite{chen2025goku},  potentially tackles such a challenging task.
However, RS HSIs are required to preserve physically reliable spectrum instead of being ``pretty pictures''.
Therefore, popular perceptual loss functions (e.g., LPIPS \cite{LPIPS} and DISTS \cite{DISTS}) or generative models are less practical for this application.
Conversely, a lightweight and explainable design is preferable to ensure the efficiency and reliability of RS applications, such as the deep unfolding-based approaches. 

In deep unfolding paradigms \cite{deepunfolding_CVPR}, a deep network is designed according to the physical interpretation of optimization criteria, resulting in a theoretically grounded architecture.
Nevertheless, numerous formulations for deep unfolding adopt a handcrafted prior (e.g., sparsity) \cite{mou2022deep,marivani2020multimodal} or an implicit deep image prior \cite{zhang2024dual}.
The first prior has less effective modeling for practical scenarios; the latter one essentially relies on complex architectural design, which is similar to those unexplainable deep models.
If the unfolded network ultimately remains data-driven, why not learn the desirable prior from the training data directly to enable a superior formulation?
To this end, we propose a lightweight PriorNet (containing only 0.05 M parameters and 2.6 G MACs) to empower the data-fitting model and regularization.
The PriorNet performs a preliminary spatial resolution unification to 5 m GSD in the multispectral domain, thereby obtaining the spatial prior image.
With the prior image, the subsequent criterion design can focus on the challenging 12-to-186 SSR, rather than simultaneously tackling two challenging tasks.
Furthermore, the PriorNet also provides a spectral prior matrix (SPM) that encodes the cross-similarity among 186 hyperspectral bands.  
Since the SSR task is highly ill-posed, we constrain the estimated HSI to maintain a comparable spectral cross-similarity with the SPM, namely, data-driven spectrum regularization.

In addition, a fundamentally different adversarial learning strategy is developed to achieve effective design.
In a conventional generative adversarial network (GAN) \cite{chan2022efficient,CODE}, the discriminator outputs a probability indicating whether the input is real or fake, thereby enforcing the generator toward a realistic output.  
However, the well-trained discriminator will be discarded after the training stage.
Due to the substantial demand for reliability in RS applications, \textit{Can the discriminator continuously supervise the model to ensure a reliable solution in the inference phase?}
Therefore, we employ the discriminator maximization term in our criteria, encouraging the final solution to fool the discriminator.
Through the Qusai-split Bregman (Qusai-SB) iteration, the HSI is progressively updated to maximize the probability output from the discriminator, yielding an effective solution.
Consequently, the explainable generator, built by unfolding the Qusai-SB optimization, can conduct adversarial learning without an explicit adversarial loss function.
Meanwhile, the discriminator, trained jointly with the generator, not only enforces a stronger generator during training but also provides continuous guidance during testing, i.e., unfolding adversarial learning (UAL).
The experiment will demonstrate that the superior performance (in PSNR, SSIM, SAM, and RMSE metrics) arises from our explainable design (over 20 times fewer parameters than the next-best Transformer) instead of increasing architectural depth or complexity.
%

%
\section{Related Work} \label{sec:Related}
\subsection{Conventional Spectral Super-Resolution}
Because of the high cost of hyperspectral imaging systems, reconstructing the HSI from its RGB counterpart has become a practical alternative solution.
Therefore, spectral reconstruction (i.e., SSR) has gained considerable attention in the computer vision and image processing fields.
Several well-known datasets, such as NUS \cite{NUS}, CAVE \cite{CAVEdata}, Harvard \cite{Harvard}, provide paired RGB images and HSIs captured by various hyperspectral cameras, thereby facilitating the development of SSR algorithms to produce the high spectral-spatial resolution HSI.  
For example, traditional algorithms, including regression-based \cite{eslahi2009recovery}, sparse coding \cite{arad2016sparse,8116687,8410422}, manifold learning \cite{jia2017rgb}, and prior-based \cite{8481553,9109715}, have been extensively investigated. 
Recently, data-driven algorithms \cite{zhang2022survey,HSGAN,MSFN,SSRNet,alvarez2017adversarial,8575293} have been widely applied to learn the mapping between RGB images and their hyperspectral counterpart. 
In the recent NTIRE spectral reconstruction challenges, the winners from 2018 to 2022, including \cite{NTIRE}, HSCNN+ \cite{HSCNN+}, adaptive weighted
attention network (AWAN) \cite{AWAN}, and multi-stage spectral-wise Transformer (MST++) \cite{MST++} are predominantly data-driven models, substantiating the superiority of data-driven approaches.
However, these methods are primarily designed for CAVE-level spectral reconstruction from RGB images (3-to-31 SSR).
When insisting on applying them to our target transformation, the computational burden may increase rapidly, whereas the performance is difficult to guarantee.
\subsection{Sentinel-2 to AVIRIS-Level Reconstruction}
To the best of our knowledge, this work is the first customized algorithm targeting AVIRIS-C level SSR (12-to-186) together with AVIRIS-NG level SRU (5 m GSD).
The most closely related literature would be the very recent ``Conversion from Sentinel-2 to AVIRIS'' (COS2A) \cite{COS2A}, which achieves 12-to-172 SSR and SRU to 10 m GSD.
COS2A first employs a lightweight deep model to learn the spectral mapping between Sentinel-2 and AVIRIS-level HSI, and then formulates the estimated HSI and the native high-resolution Sentinel-2 bands as a convex optimization-based data fusion problem, namely the spectral-spatial duality theorem \cite[Theorem 1]{COS2A}.
Nevertheless, the 5 m GSD MSIs are physically unavailable for Sentinel-2 constellations; this duality-based data fusion strategy is not applicable.
By comparison, algorithms are required to produce the high spectral-spatial resolution AVIRIS-level HSI without such high-resolution guidance.
Motivated by the above restrictions, we aim to develop an effective, lightweight, and explainable architecture to fulfill this goal.

%% file: sec/2_Method.tex
\section{The Proposed UALNet}\label{sec: Method}
This section elaborates on the design and implementation of the proposed UALNet.
Initially, Section \ref{subsec: Criterion} describes the formulation of our optimization criterion.
Section \ref{subsec: Quasi-ADMM} then derives the update implementations of the proposed criterion based on the Qusai-SB optimization framework.
Finally, Section \ref{subsec: Unfolding} constructs the proposed UALNet by unfolding the SB iteration.
%
%
\subsection{Criterion Design}\label{subsec: Criterion}
Given a multiresolution Sentinel-2 image $\bS\in\mathbb{R}^{12\times l}$, where $l$ denotes the number of pixels in the 10 m GSD bands.
Note that although the native Sentinel-2 sensors acquire bands at three different spatial resolutions (i.e., 10 m, 20 m, and 60 m GSD), the official products are copied by a factor of 4 and 36 for those 20 m bands and 60 m bands, respectively, to construct a shape-unified multispectral data.
Our objective is to transform $\bS$ into the AVIRIS-level HSI $\bA\in\mathbb{R}^{186\times L}$ with $L=4l$ (5 m GSD).

Because of the multiresolution nature of Sentinel-2 data, the spectral response function (SRF), non-uniform spatial blurring, and downsampling matrices must be considered to model the relationship between $\bS$ and $\bA$. 
However, explicitly modeling the relationship using the non-uniform spatial blurring and downsampling matrices potentially requires a vectorized representation, resulting in a large-scale matrix inversion, which is undesirable for architectural designs.
To tackle these limitations, we develop a lightweight and efficient PriorNet (see Supplementary Figure 1), denoted as $f_{\theta_p}(\cdot)$, to unify the spatial resolution to 5 m, i.e., $\bS_u=f_{\theta_p}(\bS)\in\mathbb{R}^{12\times L}$.

Accordingly, the relation between $\bS_u$ and $\bA$ can be explicitly modeled by the data-fitting (DF) term, i.e.,
\begin{align*}
     \text{DF}(\bA):=\frac{1}{2}\|\bS_u-\bD\bA\bB\|_F^2,
\end{align*}
where $\bD\in\mathbb{R}^{12\times 186}$ and $\bB\in\mathbb{R}^{L\times L}$ denote the SRF $\bD\in\mathbb{R}^{12\times 186}$ and uniform spatial blurring matrix, respectively.
Note that under a high-scale spatial SR (from $2\times$ to $12\times$) and limited data acquisition, $\bS_u$ inevitably exhibits a very slight blur.
To account for this effect, the DF term jointly considers the blurring matrix in addition to the SRF. 
In our implementation, $\bB$ is considered Gaussian blurring with a kernel size of $7$ and $\sigma:=0.7$, whereas $\bD$ is directly unfolded as learnable network parameters (see Section \ref{subsec: Unfolding}) as the SRF between Sentinel-2 and AVIRIS is unavailable.

\begin{figure}[t]
    \centerline{\includegraphics[width=0.5\textwidth]{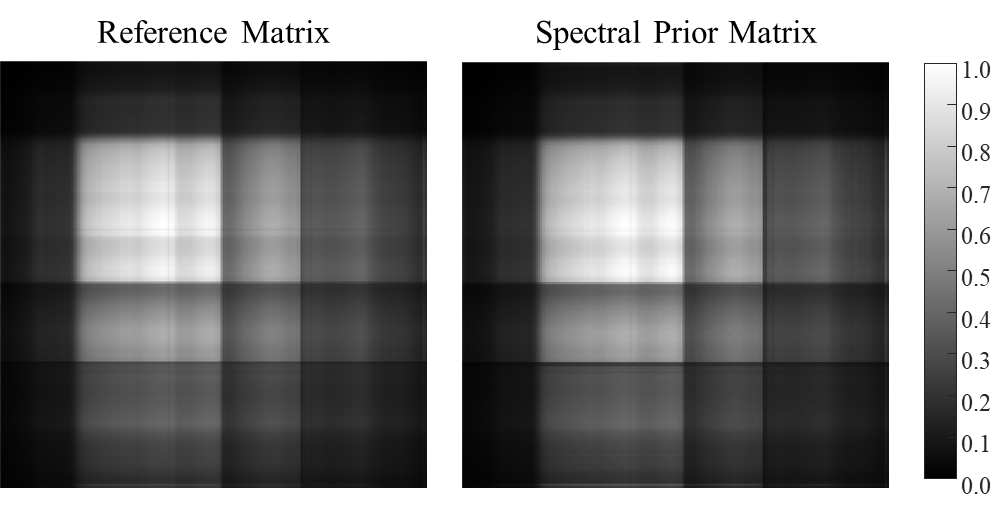}}
    \caption{Visual comparison of the reference spectral cross-similarity matrix $\bA\bA^T$ and the spectral prior matrix $\bP$ learned by the proposed PriorNet (see Supplementary Figure 1).}\label{fig: SRM}
\end{figure}

In addition, regularizations are required due to the high ill-posedness of the SSR problem.
The first is a novel discriminator-maximization regularization (DMR).
In the popular GAN paradigm, the generator competes with the discriminator by attempting to fool the discriminator, and the discriminator struggles to distinguish whether the input is real or fake.
Inspired by this, we design the DMR to encourage a more reliable reconstruction, i.e., $\frac{\lambda_1}{2}\|\textbf{1}_{186\times L}-D_{\theta_D}(\bA)\|_F^2$,
%
%
where $\textbf{1}_{M\times L}$, $D_{\theta_D}$, and $\theta_D$ denote the all-one matrix with shape $M\times L$, the discriminator function, and its parameters, respectively.
Moreover, $D_{\theta_D}(\bA)\in[0,1]^{186\times L}$ represents the probability matrix, where each entry is expected to approach 1 for real inputs and 0 for the fake ones.
The entry-wise probability may encourage $D_{\theta_D}$ to focus on unnatural regions rather than relying on specific features.
By minimizing the DMR, the desirable HSI $\bA$ is encouraged to fool the discriminator, yielding a more effective reconstruction. 
%
%
The data-driven spectrum regularization is also incorporated to preserve the spectral correlation, i.e., 
\begin{align}\label{eq: SPM}
    \frac{\lambda_2}{2}\|\bA\bA^T-\bP\|_F^2.
\end{align}
Specifically, the proposed PriorNet is also designed to approximate the SPM $\bP\approx\bA\bA^T\in\mathbb{R}^{186\times 186}$ as visualized in Figure \ref{fig: SRM}, yielding the overall regularization scheme, i.e., $\text{REG}(\bA):= \frac{\lambda_1}{2}\|\textbf{1}_{186\times L}-D_{\theta_D}(\bA)\|_F^2+\frac{\lambda_2}{2}\|\bA\bA^T-\bP\|_F^2$. 
Finally, the overall criterion can be explicitly denoted as
\begin{align}\label{eq: overall}
   \bA^{\star}=\arg\min_{\bA}~\text{DF}(\bA)+\text{REG}(\bA),
\end{align}
and the optimization strategy for solving \eqref{eq: overall} is presented in the next section.

\subsection{Quasi-Split Bregman Optimization}\label{subsec: Quasi-ADMM}
In this study, we solve \eqref{eq: overall} with an SB-inspired optimizer.
While the convergence of standard SB is only guaranteed for convex objectives with linear constraints \cite{Bregman}, our final goal is to build an explainable and reliable non-convex deep architecture.
Therefore, we employ the dual-space decouple property of SB optimization steps without pursuing the strict convexity, referred to as Qusai-SB. 
Despite the lack of provable convexity, the experiments (see Section \ref{sec: Exp}) substantially validate the effectiveness of our Qusai-SB scheme. 

To this end, we introduce an auxiliary variable $\bT\in\mathbb{R}^{186\times L}$ and hence reformulate \eqref{eq: overall} as
\begin{align}\label{eq: reformulation}
     \min_{\bA}~&\frac{1}{2}\|\bS_u-\bD\bA\bB\|_F^2  +\frac{\lambda_1}{2}\|{\bf 1}_{M\times L}-\bT\|_F^2 \notag
     \\
     &+\frac{\lambda_2}{2}\|\bA\bA^T-\bP\|_F^2 ~~~ \text{s.t.}~~~ \bT=D_{\theta_D}(\bA).
\end{align}
The augmented Lagrangian of \eqref{eq: reformulation} can be expressed as $\mathcal{L}(\bA,\bT,\bU)\!=\!\frac{1}{2}\|\bS_u\!-\!\bD\bA\bB\|_F^2\!+\!\frac{\lambda_1}{2}\|{\bf 1}_{M\times L}\!-\!\bT\|_F^2\!+\!\frac{\lambda_2}{2}\|\bA\bA^T\!-\!\bP\|_F^2\!+\!\frac{\mu}{2}\|D_{\theta_D}(\bA)\!-\!\bT\!-\!\bU\|_F^2$,  
%
%
where $\bU\in\mathbb{R}^{186\times L}$ and $\mu>0$ are the scaled dual variable and the penalty parameter, respectively.
In addition, the iterative updated steps are summarized in Algorithm \ref{alg: main_1}. 
	\begin{algorithm}[t]
		\caption{Optimization-based Algorithm for Solving \eqref{eq: reformulation}}
		\begin{algorithmic}[1]\label{alg: main_1}
            \STATE Given a Sentinel image $\bS_u$, SRF $\bD$, and a predefined discriminator $D_{\theta_D}$.
            \STATE Set $k:=0$.
            \STATE Initialize $\bA^k$ and $\bU^k$. 

            \REPEAT

            \STATE Update ${\bT}^{k+1}=\arg \min_{\bT} \mathcal{L}(\bA^k,\bT,\bU^k)$.
            
            \STATE Update ${\bA}^{k+1}=\arg \min_{\bA} \mathcal{L}(\bA,\bT^{k+1},\bU^k)$.

            \STATE Update ${\bU}^{k+1}=\bU^k-(D_{\theta_D}(\bA^{k+1})-{\bT}^{k+1})$.
            
		    \STATE $k:=k+1$.
            \UNTIL the predefined stopping criterion is met.
            \STATE {\bf Output} high spectral-spatial resolution HSI $\bA^k$.
		\end{algorithmic}
	\end{algorithm}
%
In Line 6 of Algorithm \ref{alg: main_1}, the $\bT^{k+1}$ can be updated by the following closed-form solution, i.e.,
\begin{align}\label{eq: T_k_p_1}
    {\bT}^{k+1}:={\frac{1}{\lambda_1+\mu}(\lambda_1{\bf 1}_{M\times L}+\mu\bR)},
\end{align}
where $\bR=D_{\theta_D}(\bA^{k})-\bU^{k}$.
In contrast, updating ${\bA}^{k+1}$ is relatively tricky due to the non-convexity of $D_{\theta_D}(\cdot)$ and the biquadratic term $\frac{\lambda_2}{2}\|\bA\bA^T\!-\!\bP\|_F^2$. 
Accordingly, we adopt the gradient-based implementation to obtain $\bA^{k+1}$, i.e., $\bA^{k+1}:=\bA^{k}-\gamma(\bG_1+\bG_2+\bG_3)$, where 
\begin{align}\label{eq: gradients}
&\bG_1={\bD^T}(\bD\bA^k\bB){\bB^T}-{\bD^T}\bS_u{\bB^T}, \notag
\\
&\bG_2=2\lambda_2~(\bA^k{\bA^k}^T-\bP)\bA^k, \notag
\\
&\bG_3=\mu~\mathcal{J}_{\bA^k}^T\left(D_{\theta_D}(\bA^{k})-\bT^{k+1}-\bU^{k}\right),
\end{align}
with $\mathcal{J}_{\bA^k}=\partial D_{\theta_D}(\bA^k)/\partial \bA^k$.
Besides, $\gamma>0$ presents the learning step size.
Thus far, the implementations of this Qusai-SB optimization have been completed.

\subsection{Adversarial Unfolding}\label{subsec: Unfolding}
\begin{figure*}[t]
    \centerline{\includegraphics[width=1\textwidth]{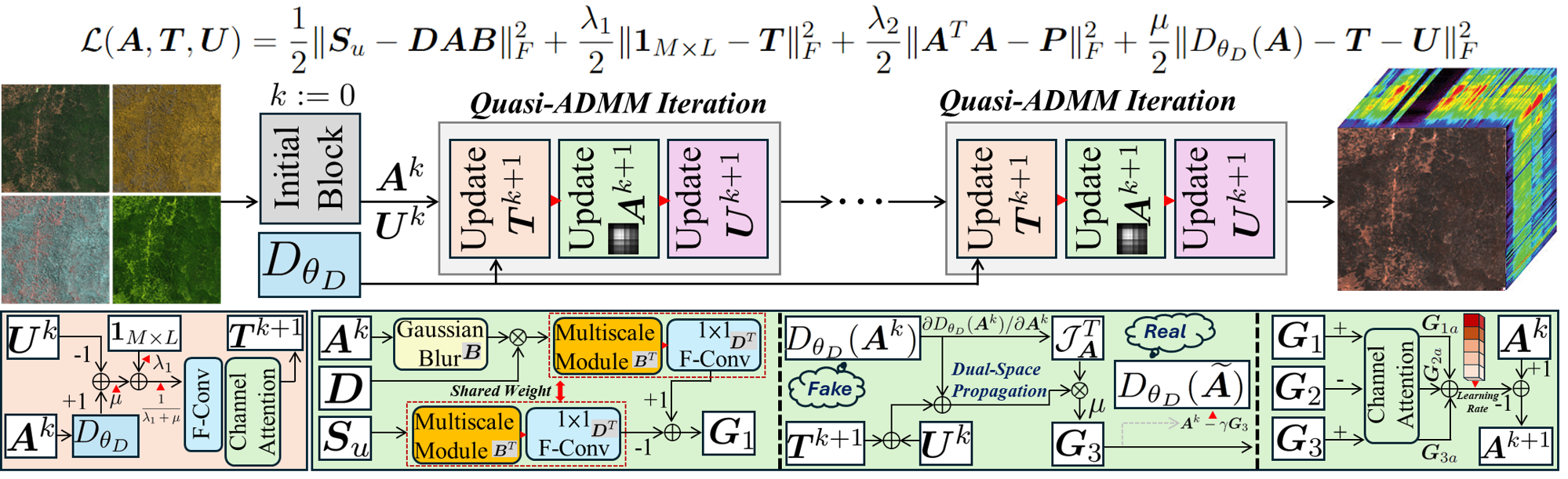}}
    \caption{The schematic pipeline of the proposed UALNet for the challenging Sentinel-2 $\bS$ to AVIRIS-level HSI $\bA$ transformation.
    To fulfill this goal, we first develop an efficient PriorNet (see Supplementary Figure 1) to provide the 5 m GSD spatial prior image $\bS_u$ from the target multiresolution MSI $\bS$, together with a spectral prior matrix $\bP\approx\bA\bA^T$ that encodes the spectral cross-correlations.
    Subsequently, the adversarial learning process is formulated as a discriminator-maximization term (see Section \ref{subsec: Criterion}) and is solved by the Qusai-SB optimization framework (see Section \ref{subsec: Quasi-ADMM}).
    By unfolding the iterative Qusai-SB procedure (see Section \ref{subsec: Unfolding}), the proposed unfolding adversarial network (UALNet) can achieve an explainable architecture.
    Due to the space limitation, the details of the sub-modules are illustrated in the Supplementary Material.
    %
    %
    }\label{fig: Adversarial_Unfolding}
\end{figure*}
The explainable architecture constructed by Algorithm \ref{alg: main_1} is illustrated in Figure \ref{fig: Adversarial_Unfolding}.
We first elaborate on the unfolding module for updating $\bT^{k+1}$ (see the orange block in Figure \ref{fig: Adversarial_Unfolding}).
Although its closed-form solution is derived in \eqref{eq: T_k_p_1}, we further incorporate a full-convolutional layer followed by a channel attention module to enhance the adaptability of the proposed UALNet.
When updating the $\bT^{1}$ ($k:=0$), the initialization of $\bA^0$ and $\bU^0$ plays a crucial role due to the non-convexity of the network.
For the $\bA^0$, the simplified residual-based network \cite{HSCNN+} with depth of 10 (while it requires a depth of 30 for CAVE-level reconstruction and 80 for this challenging task) is employed for initialization, whereas $\bU^0$ is empirically initialized as an all-zeros matrix to ensure a stable optimization.   
Subsequently, we detail the unfolding module for updating $\bA^{k+1}$ (see the green block of Figure \ref{fig: Adversarial_Unfolding}).
The relation between $\bA^{k}$ and $\bA^{k+1}$ [see \eqref{eq: gradients}] can be interpreted as a residual learning process, where $\bA^{k}$ acts as the main component and $(\bG_1+\bG_2+\bG_3)$ refers to residual features.
In implementing $\bG_1$, the SRF $\bD$ is unfolded as the trainable network parameters share-weighted across all unfolding modules, while $\bB$ is the Gaussian blur with kernel size of $7$ and $\sigma:=0.7$.
In addition, the corresponding transpose operators, $\bD^T$ and $\bB^T$, are treated as the spectral upsampling (12-to-186 linear projection) and adjoint blurring process, respectively.
Based on these intuitions, $\bD^T$ is implemented using a $1\times 1$ full-convolutional layer, whereas the $\bB^T$ is realized by the multiscale module [see Supplementary Figure 2(a)] since the spatial operator is relatively sophisticated and may benefit from various receptive fields.
Note that the $\bG_1$ represents the variation between $\bD\bA^k\bB$ and $\bS_u$ after transposed operators, thus the modules to implement $\bD^T$ and $\bB^T$ are share-weighted for $\bD\bA^k\bB$ and $\bS_u$ within a single module.

Subsequently, we describe the module design for computing $\bG_2$.
In the second line of \eqref{eq: gradients}, the discrepancy between the estimated spectral correlation $\bA^k{\bA^k}^T$ and SPM $\bP$ affects $\bA^k$ along the spectral dimension, resembling the spectral-self-attention \cite{MST++}. 
To enhance the representation and capture informative structures, we incorporate the feedforward block $F(\cdot)$ composed of a single fully-connected layer followed by a GeLU activation \cite{GELUs}, resulting in the implementation $\bG_2=2\lambda_2(F(\bA^k{\bA^k}^T)-F(\bP))\bA^k$.
As for the $\bG_3$, we directly follow the last line of \eqref{eq: gradients} to preserve the discriminative guidance provided by the discriminator.
The implementation is realized using Pytorch \cite{paszke2019pytorch} automatic differentiation, while the computational cost (measured in MACs) is approximated as twice that of a single forward process.
Given these informative features $\bG_1$, $\bG_2$, and $\bG_3$, we adopt independent channel attentions to adaptively adjust the different channel-wise learning steps, yielding the attention features, $\bG_{1a}$, $\bG_{2a}$, and $\bG_{3a}$, as shown in Figure \ref{fig: Adversarial_Unfolding}.
Consequently, we have the final update $\bA^{k+1}:=\bA^{k}+\gamma(\bG_{1a}+\bG_{2a}+\bG_{3a})$ with learnable global step $\gamma$.
This completes the overall design of the UALNet.

For the loss function design, the discriminator $D_{\theta_D}$ outputs element-wise real/fake probabilities, i.e., $D_{\theta_D}(\bA)\in [0,1]^{186\times L}$, to distinguish whether the input is real or fake.
Given the reference HSI $\bA$ and the UALNet estimation $\widehat{\bA}$, the loss function of $D_{\theta_D}$ can be explicitly written as $ \mathcal{L}_D:=-[\log(p_{\text{r}})+\log(1-p_{\text{f}})]$,
%
%
where $p_{\text{r}}=\frac{1}{186L}\sum_{i,j}[D_{\theta_D}(\bA)]_{i,j}$ and $p_{\text{f}}=\frac{1}{186L}\sum_{i,j}[D_{\theta_D}(\widehat{\bA})]_{i,j}$.
For the loss function of UALNet, we adopt the spectrally adaptive $\ell$-1 loss, i.e., 
\begin{align}
    \mathcal{L}_{\text{G}}:=\frac{1}{186L}\sum_{i=1}^{186} \sum_{j=1}^{L}\alpha_j|[\bA]_{i,j}-[\widehat{\bA}]_{i,j}|,
\end{align}
where $\alpha_j$ denotes the spectral angle mapper (SAM) \cite{HyperQUEEN} between $[\bA]_{:,j}$ and $[\widehat{\bA}]_{:,j}$.
Because the architectural design of the proposed UALNet has embedded the adversarial learning process via DMR [see \eqref{eq: reformulation} and Algorithm \ref{alg: main_1}].
Therefore, UALNet does not require the adversarial loss function.
In addition, Section \ref{subsec: Settings} presents the training implementations.
%

%

%% file: sec/3_Experiment.tex
\section{Experiment}\label{sec: Exp}
This section describes the experiments to demonstrate the effectiveness of our framework.
Section \ref{subsec: Dataset} provides the details of the data preparation.
The corresponding training settings are described in Section \ref{subsec: Settings}.
Subsequently, Section \ref{subsec: Evaluations} reports both the qualitative and quantitative results.
Finally, a comprehensive ablation study and a case study are discussed in Section \ref{subsec: Ablation} and Section \ref{subsec: Case}, respectively.
\subsection{Dataset}\label{subsec: Dataset}
\begin{figure}[t]
    \centerline{\includegraphics[width=0.5\textwidth]{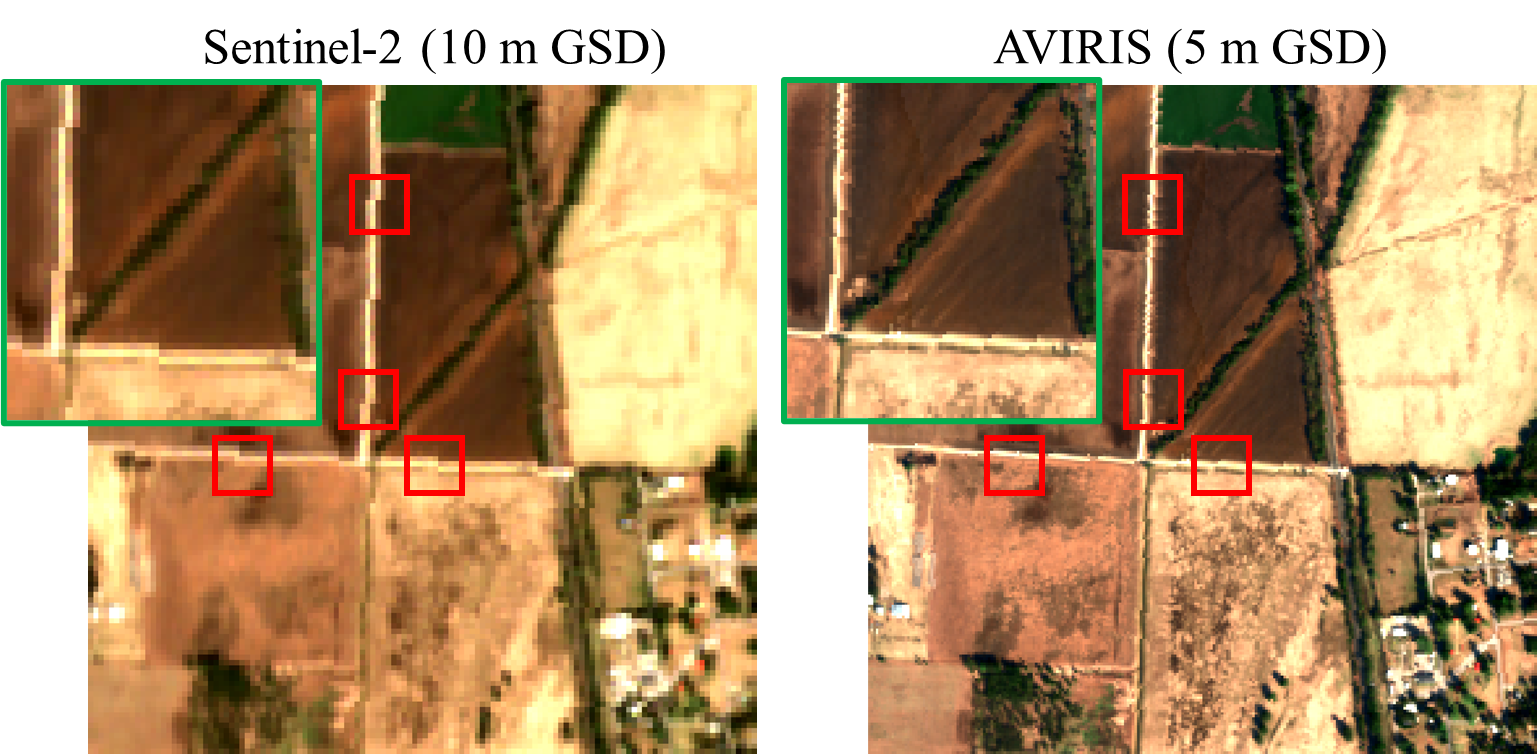}}
    \caption{Comparison of spatial calibration between the 10 m GSD Sentinel-2 MSI and the 5 m GSD AVIRIS HSI in true-color composition.}\label{fig: misaligment}
\end{figure}
In this study, we made extensive efforts to collect the spatially and temporally aligned Sentinel-2 MSI and AVIRIS-NG HSI for the real-world applicability of our UALNet.
Nevertheless, such real data pairs are limited, and no public dataset is available.
Currently, we have prepared 415 AVIRIS-NG HSIs \cite{AVIRISNG} with a spatial size of $256\times 256$, whereas only a subset of the paired Sentinel-2 MSIs are collected.
On the other hand, the flying trajectories of Sentinel-2 and AVIRIS-NG are not naturally aligned; thus, spatial calibrations, including rotation, pixel-level registration, are the necessary preprocessing steps.
Given the high resolution of AVIRIS-NG, even slight pixel-level deviation (see Figure \ref{fig: misaligment}) would bias the model learning process and result in suboptimal estimations. 
We are currently working on the spatial calibrations as well.
Consequently, 365, 20, and 30 data pairs (Sentinel-2 and AVIRIS-NG) simulated using AVIRIS-NG HSIs are adopted for training, testing, and validation, respectively.

The experimental protocol for simulation data is described as follows.
Initially, we remove the water vapor bands, i.e., band 1, bands 195-211, and bands 281-315, from the original AVIRIS-NG 425-band HSI \cite{AVIRISNG}.
The remaining 372 bands are resampled in ENVI \cite{ENVI} to a GSD of 5 m, and then a spectral downsampling by a factor of 2 to obtain 186 high-quality bands.
For the subsequent multiresolution simulation, these 186-band HSIs are further resized from $256\times 256$ to $252 \times 252$ in MATLAB, resulting in the reference HSIs $\bA$.
To simulate Sentinel-2 MSIs, we first produce the uniform spatial resolution (5 m GSD), 12-band MSI $\bS_u=\bD\bA$.  
Although the public SRF $\bD\in\mathbb{R}^{12\times 186}$ between $\bA$ and $\bS_u$ is not available, we approximate it by averaging the closest subset bands of $\bA$ based on each Sentinel-2's bandwidth and central wavelength \cite{SSSS,S2data}.
With the $\bS_u$, the spatial modeling in \cite{SSSS} is extended to obtain the simulated Sentinel-2 MSIs $\bS$.
Specifically, we apply the circular Gaussian blurring with downsampling factors of 12, 2, 2, 2, 4, 4, 4, 2, 4, 12, 4, and 4 on the band 1-12 of $\bS_u$.
Subsequently, we spatially downsample each blurred band by the same factors, thereby obtaining the 60 m (B1, B10), 10 m (B2, B3, B4, B8), and 20 m (B5, B6, B7, B9, B11, B12) GSD bands.
To form a typical matrix, the 60/20 m bands are copied 36/4 times to have a consistent spatial dimension as the 10 m bands, which is the same step as the real Sentinel-2 product downloaded from the official website.
Finally, 365, 20, and 30 data pairs are randomly split to form the training, testing, and validation sets, respectively.
\begin{table*}[t]
\footnotesize
\centering
\caption{Performance and efficiency comparisons between the proposed UALNet and the SOTA algorithms, where the red and blue boldfaced numbers indicate the best and the next-best quantitative metrics (reported in PSNR, SAM, SSIM, and RMSE), respectively.}\label{tab: performance}
\begin{tabular}{c|cccccccccc} 
\hline \hline
Methods   & HSCNN+  & HRNet & Restormer & MST & MST++ & SSRNet & MSFN   &  SPECAT  & COS2A & UALNet       
\\ \hline
PSNR$~\!(\uparrow)$ & 30.0771 & 28.8220 & 29.4884 & 29.9067 & 31.6803 & 25.0876 & 29.3009 & 31.5218 & {\bf \blue 32.2976} & {\bf \red 32.5986} %
\\
 SAM$~\!(\downarrow)$ & 3.1940 & 4.2748 & 3.5788 & 3.3406 & 2.7452 & 13.9854 & 3.6518 & {\bf \blue 2.6984} & 3.3148 & {\bf \red 2.4869}
\\
 SSIM$~\!(\uparrow)$ & 0.8966 & 0.8856 & 0.8930 & 0.8952 & {\bf \blue 0.9187} & 0.8381 & 0.8927 & 0.9177 & 0.8897 & {\bf \red 0.9214}%
\\
RMSE$~\!(\downarrow)$  & 0.0177 & 0.0195 & 0.0176 & 0.0170 & {\bf \blue 0.0148} & 0.0494 & 0.0179 & 0.0149 & 0.0178 & {\bf \red 0.0145}
\\
\hline
Params (M)  & 20.9889 & 34.8269 & 15.2310 & 258.8675 & 56.6967  & 12.5077 & 130.2586 & 15.3738 & 4.5914 & 1.7554
\\
MACs (G)  &  1053.1404 & 133.7938 & 94.6391 & 972.3171 & 804.0262  & 793.6337  &1680.6228  & 512.1194 & 230.1611 & 120.1429
\\ \hline \hline
              
\end{tabular}
\end{table*}
\begin{figure*}[t]
    \centerline{\includegraphics[width=1\textwidth]{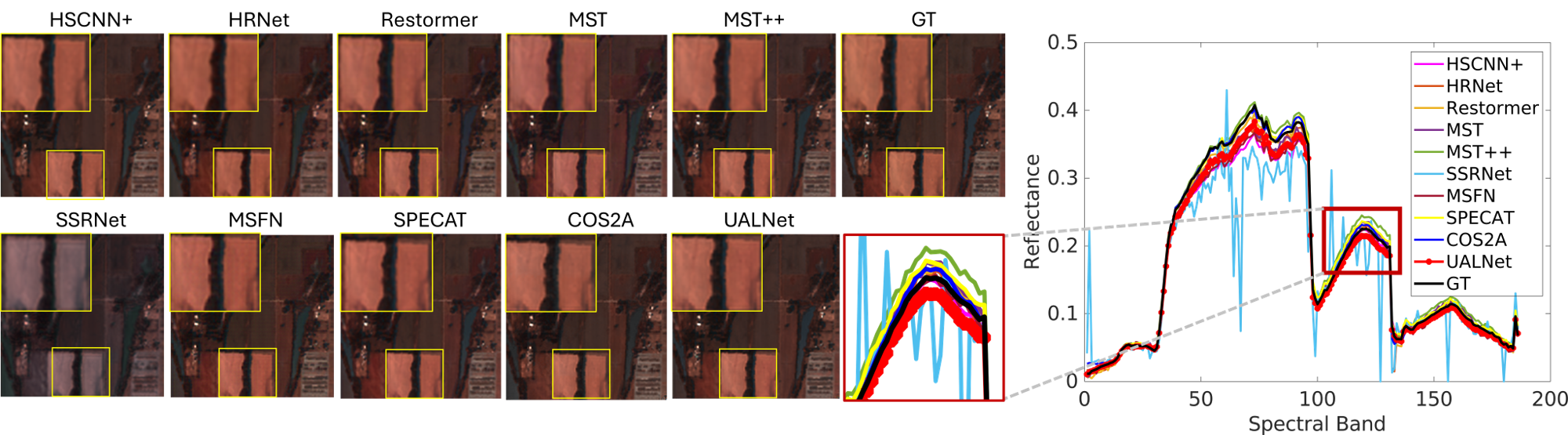}}
    \caption{Qualitative comparisons between the estimated results and the corresponding GT, shown in true-color compositions (left) and the spectral signatures (right).
    The ROI is located near Okmulgee County, Eastern Oklahoma, USA, and was captured on Oct. 27, 2019.}\label{fig: exp_1}
\end{figure*}
\begin{figure*}[t]
    \centerline{\includegraphics[width=1\textwidth]{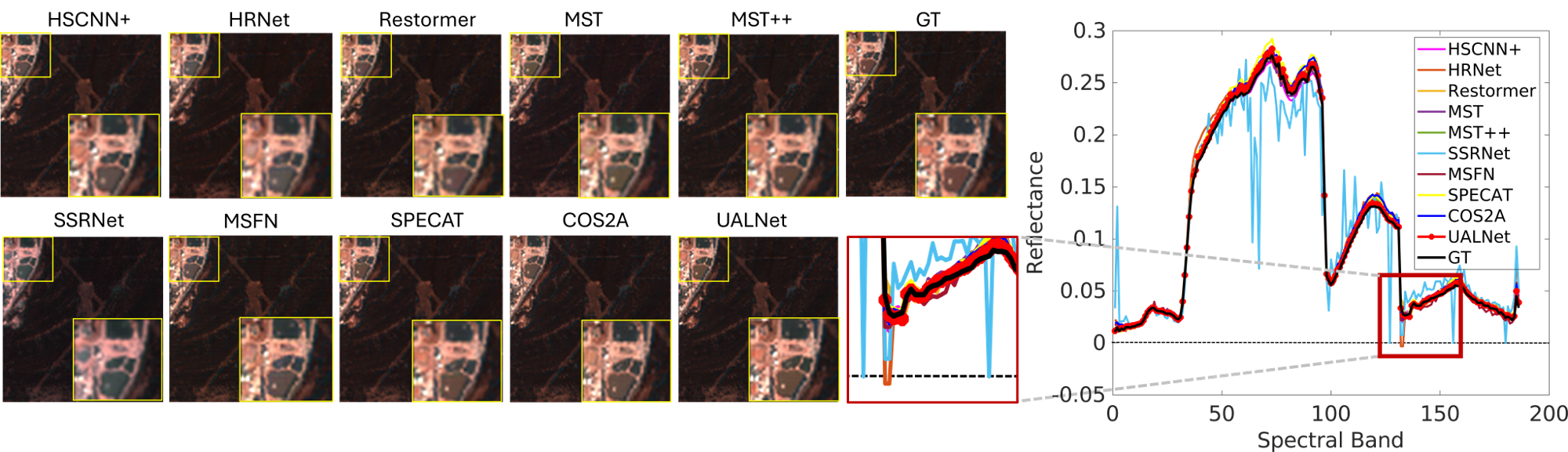}}
    \caption{Qualitative comparisons between the estimated results and the corresponding GT, shown in true-color compositions (left) and the spectral signatures (right).
    The ROI is located near Garvin County, Southern Oklahoma, USA, and was captured on Oct. 27, 2019.
    }\label{fig: exp_2}
\end{figure*}
\subsection{Experimental Setting}\label{subsec: Settings}
In our training implementation, we first pretrain the PriorNet on $252\times252$ image pairs using Adam optimizer $(\beta_1,\beta_2):=(0.9,0.999)$ with batch size of 5.
The initial learning rate is set to 5E-4 and scheduled by cosine annealing with $\eta_{\text{min}}:=$5E-5.
The number of quasi-SB iterations, $\lambda_1$, $\lambda_2$, and $\mu$ is set to 2, 5E-4, 5E-1, and 5E-2, respectively.
The UALNet and discriminator are trained using separated Adam optimizers with $(\beta_1,\beta_2):=(0.9,0.999)$ for a total of 600 epochs.
During the first 250 epochs, the discriminator and UALNet are optimized for 2 epochs in turn, using $64\times 64$ patches.
The initial learning rates are set to 5E-4 for UALNet and 1E-5 for the discriminator, which are all scheduled by cosine annealing. 
After that, we freeze the discriminator as it is strong enough, while UALNet is trained along with a learning rate of 8E-5 and $240\times 240$ patches.

We compare the proposed UALNet with numerous reconstruction-based algorithms, including HSCNN+ \cite{HSCNN+}, HRNet \cite{HRNet}, Restormer \cite{Restormer}, MST \cite{MST}, MST++ \cite{MST++}, SSRNet \cite{SSRNet}, MSFN \cite{MSFN}, SPECAT \cite{SPECAT}, and COS2A \cite{COS2A}, with appropriate modifications, such as spatial interpolation, input/output channel, or additional model depth.
Because of most benchmarks are designed for CAVE-level spectral reconstruction; therefore, their computational complexity and parameters inevitably increase exponentially (see Table \ref{tab: performance}) when extending to the Sentinel-2 to AVIRIS-level SSR task. 
Even compared to these more complex architectures, the experiments (to be described in the next section) still demonstrate that the proposed UALNet achieves superior quantitative performance due to the customized and efficient design.

\subsection{Qualitative and Quantitative Evaluations}\label{subsec: Evaluations}
The qualitative results are presented in Figure \ref{fig: exp_1} and Figure \ref{fig: exp_2}, with spectral and spatial zoomed-in figures to facilitate detailed comparisons.
As shown in Figure \ref{fig: exp_1}, most benchmarks can achieve visually acceptable results in the true-color composition, except for SSRNet.
In this challenging spectral-spatial reconstruction task, SSRNet potentially focuses more on spatial details while failing to preserve the spectral fidelity, resulting in a substantial color distortion.
In contrast, when we take a close look at the result of MSFN, it suffers from a distinct grid artifact that may be due to the patch-based attention mechanism.  
Although HSCNN+, Restormer, MST, and MST++ have visually promising results, a slightly blurred effect can still be observed.
Conversely, the proposed UALNet achieves superior spatial details with less blurring effect.
Moreover, the spectral signature comparison further substantiates that SSRNet achieves less effective spectral reconstruction, consistent with the true-color composition.
While the proposed UALNet accomplishes a more effective spectral reconstruction than the benchmarks
On the other hand, noticeable color distortion can be visually observed from the results of HRNet and SSRNet, as demonstrated in Figure \ref{fig: exp_2}.
We also notice that some benchmarks produce a negative spectral reflection, which is physically impossible in RS images.
Unlike conventional computer vision tasks, these ``fake'' artifacts should be strictly prevented in RS applications.
By comparison, the proposed UALNet benefits from continuous guidance from the discriminator, yielding more effective and reliable reconstructions.

To ensure a fair evaluation, the quantitative performance, Params, and MACs are further summarized in Table \ref{tab: performance}.
The explicit definitions of the quantitative metrics can be found in \cite{CODE} and \cite{HyperQUEEN}.
The Params and MACs reported for the proposed UALNet include the PriorNet, the discriminator, and the main architecture. 
The Params and MACs reported for COS2A solely the deep part. 
%
%
All MACs are measured on the $252\times 252$ inputs.
According to Table \ref{tab: performance}, most benchmarks require much higher memory consumption and computational complexity when modified to fulfill this task.
For instance, the next-best MST++ requires 56.65 M Params and 804 G MACs for this task. 
In contrast, our UALNet achieves a superior performance across all metrics, while having significantly cheaper computational requirements, i.e., 3\% Params and 15\% MACs compared to MST++.

\subsection{Ablation Study}\label{subsec: Ablation}
\begin{table}[t]
\footnotesize
\centering
\caption{Ablation study on the effectiveness of data-driven priors learned from the proposed PriorNet.}\label{tab: Ablation}
\begin{tabular}{c|cccc} 
\hline \hline
Methods   & PSNR$~\!(\uparrow)$  & SAM$~\!(\downarrow)$ & SSIM$~\!(\uparrow)$ & RMSE$~\!(\downarrow)$ 
\\ \hline
w  PriorNet & {\bf 32.5986} & {\bf 2.4869} & {\bf 0.9214} & {\bf 0.0145} 
\\
w/o  PriorNet & 30.4953 & 2.7711 & 0.9017 & 0.0168 
\\ \hline \hline
              
\end{tabular}
\end{table}
This ablation demonstrates the effectiveness of PriorNet by replacing $\bS_u$ with $2\times$ bicubic upsample of $\bS$ and removing spectrum regularization [see \eqref{eq: SPM}]; while retraining the UALNet in the same training setting (see Section \ref{subsec: Settings}).
As reported in Table \ref{tab: Ablation}, the UALNet with data-driven priors learned from PriorNet consistently outperforms the one without PriorNet across all quantitative metrics, validating the necessity and effectiveness of the proposed PriorNet.
Notice that the PriorNet requires only 0.05M parameters (3\% of overall UALNet) and 2.6 G MACs (2\% of overall UALNet); such a cheap computational cost is significantly worthwhile to yield a superior reconstruction.
\subsection{Case Study on Real Sentinel-2 Data}\label{subsec: Case}
\begin{figure}[t]
    \centerline{\includegraphics[width=0.5\textwidth]{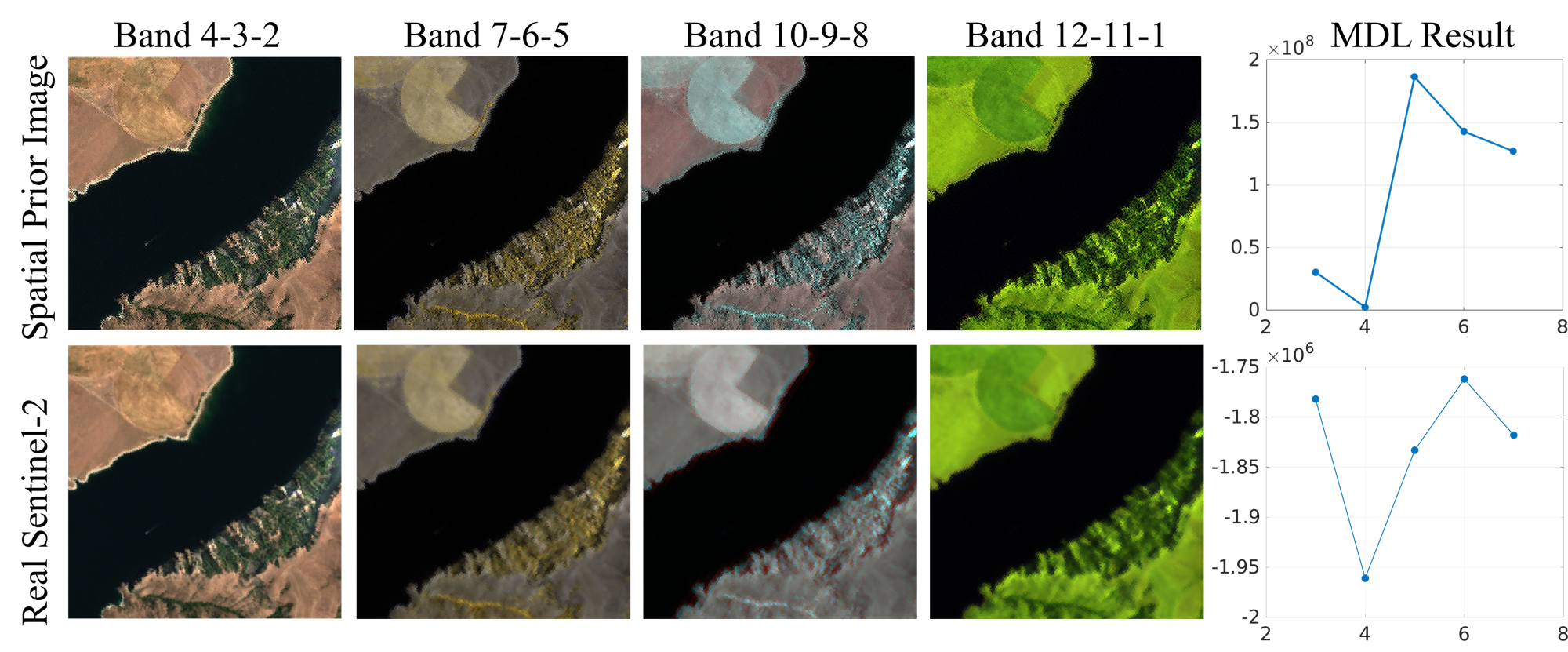}}
    \caption{Qualitative comparisons between the Prior images and the real Sentinel-2 images, and the model-order selection (the horizontal axis is the number of sources, the vertical axis indicates the code length) using the minimum description length (MDL) \cite{MDL}.}\label{fig: prior_img}
\end{figure}
%
%
This section presents a real Sentinel-2 validation.
We first show that our PriorNet also achieves a promising SRU for the real data, and preserves the reliable multispectral structure under the minimum description length (MDL) criterion \cite{MDL} (see Figure \ref{fig: prior_img}).
Moreover, we conduct an unmixing-based evaluation on both the Sentinel-2 image and the HSI reconstructed by UALNet; the resulting abundance maps indicate a reliable hyperspectral structure (see Figure \ref{fig: unmixing_comparison}).
Please find more analyses in the Supplemental Material.

%% file: sec/4_Conclusion.tex
\section{Conclusion}\label{sec: Conclusion}
This study, for the first time, achieves the Sentinel-2 to AVIRIS-level transformation at 5 m GSD by developing a novel UALNet.
Built upon a customized quasi-SB optimization scheme, the proposed UALNet embeds the adversarial learning process into its architecture, enabling the discriminator to continuously guide the inference phase.
Both qualitative and quantitative comparisons exhibit that our UALNet surpasses the next-best Transformer while requiring only its 3\% Params and 15\% MACs.
The real Sentinel-2 case study further substantiates the effectiveness of our framework. 

\section{Acknowledgment}
This study was partly supported by the Emerging Young Scholar Program (namely, the 2030 Cross-Generation Young Scholars Program) of the National Science and Technology Council (NSTC), Taiwan, under Grant NSTC 114-2628-E-006-002.
We thank the National Center for Theoretical Sciences (NCTS) and the National Center for High-performance Computing (NCHC) for providing the computing resources.

%% file: sec/X_Supple.tex
\clearpage
\setcounter{page}{1}
\maketitlesupplementary

\section{The Proposed PriorNet}
\label{sec: PriorNet}
\begin{figure*}[t]
    \centerline{\includegraphics[width=1\textwidth]{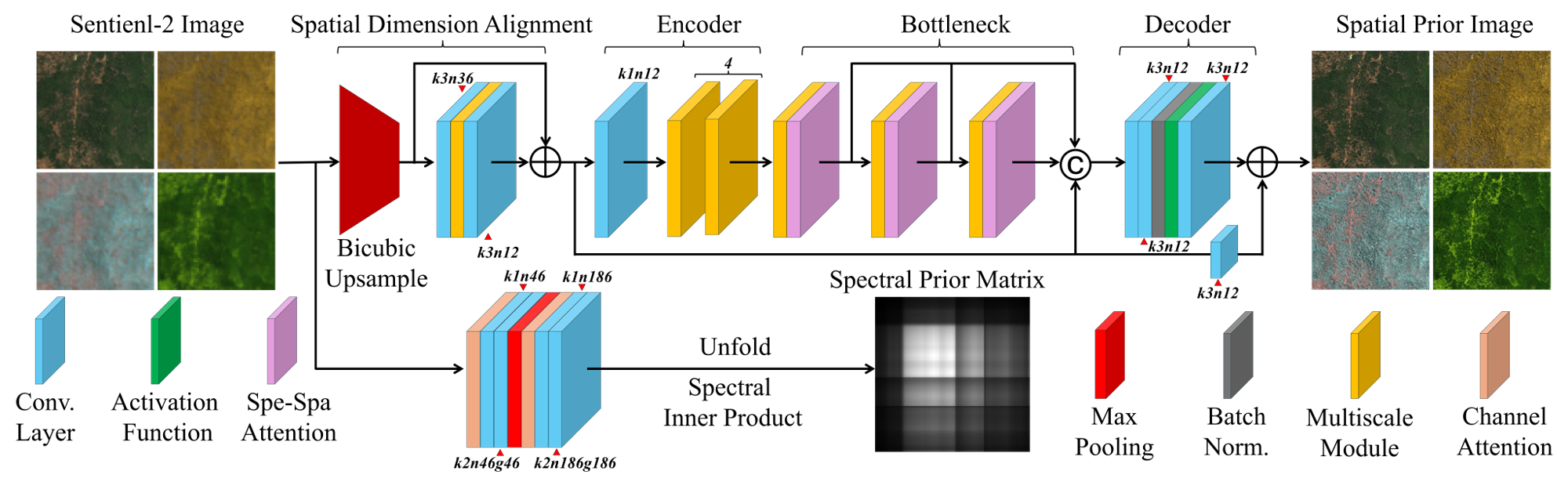}}
    \caption{Overall pipeline of the proposed lightweight PriorNet, where the notation ``k$a$-n$b$-g$c$'' denotes a 2D convolution with a kernel size of $a$, output channel of $b$, and group number of $c$ (a full convolution would not exhibit a group number additionally).
    The architectural details of Multiscale Module, Channel Attention, and Spe-Spa Attention are illustrated in Figure \ref{fig: Multiscale}(a), Figure \ref{fig: Multiscale}(b), and Figure \ref{fig: Multiscale}(c), respectively.
    In our design, the fist branch of PriorNet perform a spatial resolution unification (SRU) to obtain the high and uniform resolution (5 m GSD) prior image $\bS_u\in\mathbb{R}^{12\times L}$ from the input multiresolution (containing 60 m, 20 m, and 10 m GSD) Sentinel-2 image $\bS\in\mathbb{R}^{12\times l}$, where $L=4l$.
    The second branch provides a spectral prior matrix $\bP\in\mathbb{R}^{186\times 186}$ that encodes the spectral cross-similarity of the target AVIRIS-level HSI $\bA\in\mathbb{R}^{186\times L}$, i.e., $\bP\approx\bA\bA^T$.
    The introduction of the architectural design is presented in Section \ref{sec: PriorNet}, and Section 3 of the main article describes the role of PriorNet within the overall UALNet.
    }\label{fig: priornet}
\end{figure*}
\begin{figure}[t]
    \centerline{\includegraphics[width=0.45\textwidth]{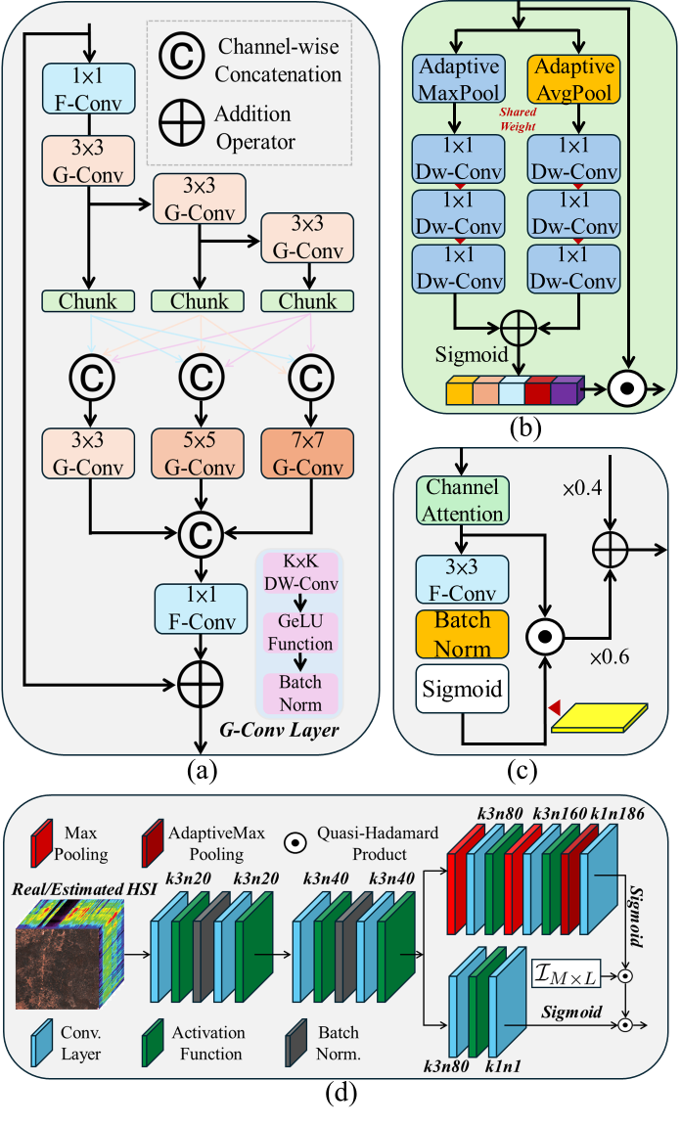}}
    \caption{Schematic diagrams of the network architectures, including (a) Multiscale Module, (b) Channel Attention, (c) Spe-Spa Attention, and (d) Discriminator.
    }\label{fig: Multiscale}
\end{figure}
The notation usages in the following description are consistent with the main article (see Section 3 of the main article).
The overall pipeline of the proposed lightweight PriorNet is shown in Figure \ref{fig: priornet}.
PriorNet comprises two branches for a) performing the spatial resolution unification (SRU) to 5 m GSD, and b) providing the spectral prior matrix $\bP$ to empower the spectrum regularization [see Eq. (1) of the main article].
Given a multiresolution Sentinel-2 data $\bS$, the first branch initially performs a 2$\times$ bicubic upsampling, followed by a short residual convolution process (composed of two convolutions and a multiscale module).
This residual block performs primary spatial dimension alignment (SDA) for the subsequent encoder.  
A single convolution then projects the SDA features to a latent space, and four cascaded multiscale modules are applied for feature encoding.
After the encoder, the bottleneck, composed of three multiscale modules and spe-spa attentions, contributes to the main signal processing.
The bottlenecks not only process the encoded features sequentially but also provide informative multi-depth features for the following fusion-based decoder.
We empirically observe that the multi-depth features significantly contribute to the performance of the output spatial prior image.
Finally, the fusion-based decoder aggregates these multi-depth features with a long-range residual connection to output the final 5 m GSD spatial prior image $\bS_u$.

In the second branch, we design the PriorNet to trade the spatial structure for precise cross-similarity of the target hyperspectral bands.
In detail, an inevitable challenge in spectral super-resolution (SSR) is to preserve both spatial and spectral structure at the same time.
Therefore, the second branch outputs a spatially compressed matrix $\widetilde{\bA}\in\mathbb{R}^{186\times \frac{l}{64}}$, which preserves the similar corss-similarity patterns as $\bA$, i.e., $\bA\bA^T\approx\widetilde{\bA}\widetilde{\bA}^T=\bP$.
Since we do not strictly demand the spatial details of $\widetilde{\bA}$, an efficient design is sufficient to yield a promising approximation, $\bP\approx\bA\bA^T$.
Then, the spectral prior matrix $\bP$ can be used to guide the overall UALNet, namely, data-driven spectrum regularization (see Section 3.1 of the main article).
To this end, channel attention is first applied to the input Sentinel-2 data to adjust the energy of each spectral band.
Then, the first convolution is adopted for the stage-one spectral upsampling (12-to-46), while the second one is for a trainable 2$\times$ spatial downsampling.
In addition, a 2$\times$ maxpooling layer with channel attention is further incorporated.
Consequently, we use two other convolutions for the stage-two spectral upsampling (46-to-186) and trainable 2$\times$ spatial downsampling to output $\widetilde{\bA}$.

For the loss function design, we first adopt the smooth $\ell$-1 loss \cite{RCNN} between the reference $\bS_u$ and the estimated result from PriorNet $\widehat{\bS_u}$ as the data-fitting loss, i.e., $\mathcal{L}_1:=\text{SmoothL1}(\bS_u,\widehat{\bS_u})$.
In addition, conventional distance-based loss functions (e.g., MSE and $\ell$-1 loss) encourage low-frequency components \cite{LPIPS}, failing to obtain fine spatial details for $\widehat{\bS_u}$.
Accordingly, the frequency loss function is adopted to resolve this issue, i.e., $\mathcal{L}_2:=\text{SmoothL1}(|\mathcal{F}(\bS_u)|,|\mathcal{F}(\widehat{\bS_u}|)$, where $\mathcal{F}(\cdot)$ indicates the 2D Fourier transform, and $|\cdot|$ is the absolute value operator.
Moreover, we preserve the spectral fidelity using the spectral angle mapper (SAM) loss, i.e., $\mathcal{L}_3:=\text{SAM}(\bS_u,\widehat{\bS_u})$.
Consequently, we also use smooth $\ell$-1 loss for the spectral prior matrix, $\mathcal{L}_4:=\text{SmoothL1}(\bA\bA^T,\bP)$.
The overall loss  function can be explicitly written as 
\begin{align*}
    \mathcal{L}_1+\lambda_1\mathcal{L}_2+\lambda_2\mathcal{L}_3+\lambda_3\mathcal{L}_4,
\end{align*}
where $\lambda_1:=$2.5E-3, $\lambda_2:=$2.5E-3, and $\lambda_3:=$5E-4.
Thus far, the introduction of PriorNet has been completed, and we refer readers to Section 4.2 of the main article for the training settings.
\begin{figure}[t]
    \centerline{\includegraphics[width=0.55\textwidth]{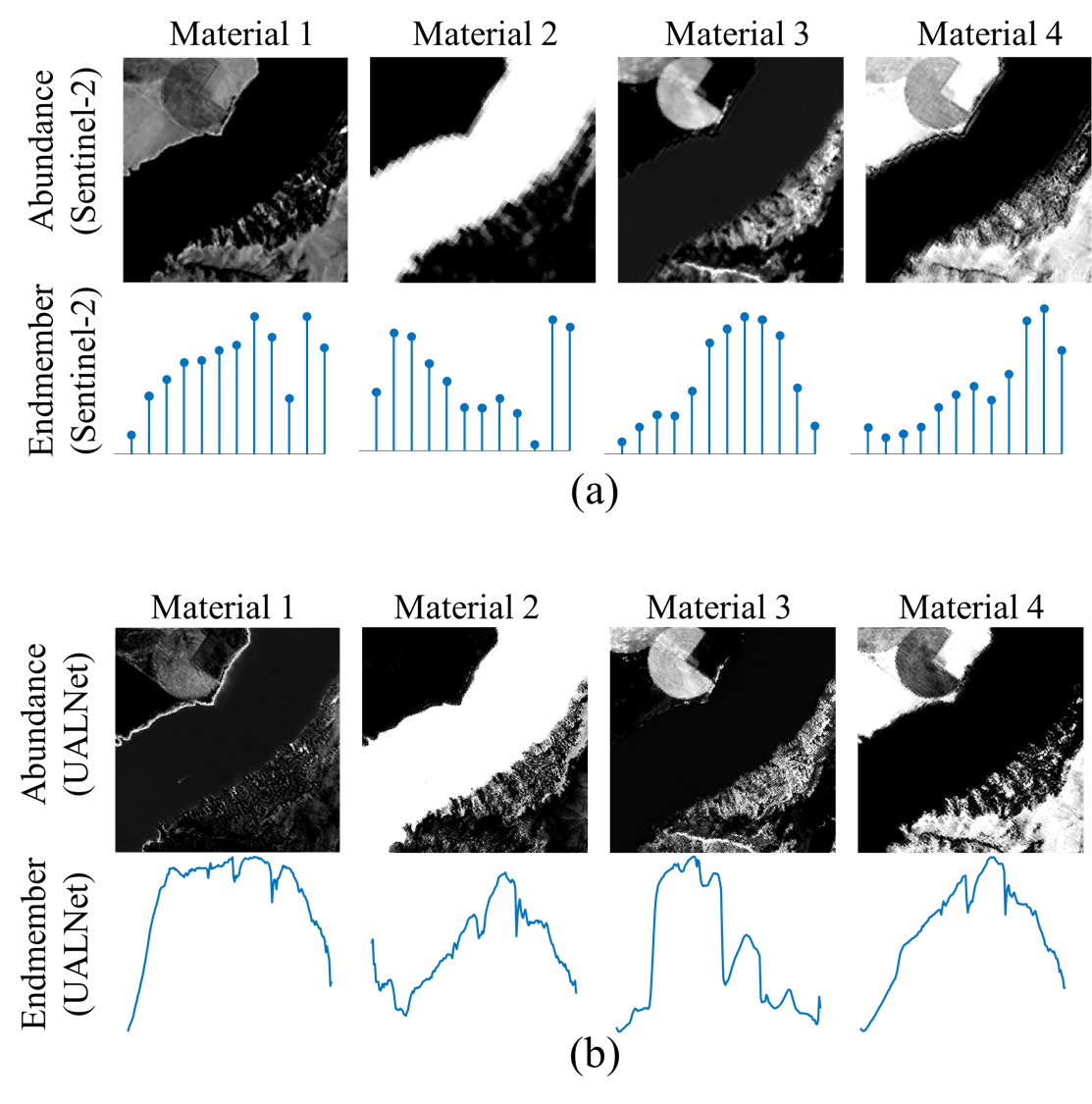}}
    \caption{Qualitative comparisons of the abundance maps and endmembers recovered from (a) Real Sentinel-2 MSI and (b) the corresponding AVIRIS-level reconstruction from the proposed UALNet.}\label{fig: unmixing_comparison}
\end{figure}
\section{Analysis for Real Sentinel-2 Case Study}
\label{sec: Case}
This section provides a detailed analysis to support the conclusion of Section 4.5 of the main article.
In this case study, we employ real Sentinel-2 data as the input of our pretrained UALNet and then reconstruct the corresponding AVIRIS-level HSI.
Nevertheless, as remarked in Section 4.1 of the main article, paired real Sentinel-2 data and AVIRIS-NG HSIs are hardly collected.
Therefore, beyond simulated quantitative evaluations, we conduct an unmixing-based evaluation using real Sentinel-2 input to demonstrate the effectiveness and practicality of the proposed framework in real-world applications.

Unmixing [i.e., blind source separation for remote sensing (RS) imagery] is a critical image processing technology for RS material identification \cite{HyperCSI,zeng2024unmixing}.
As typical RS images often suffer from the mixed-pixel phenomenon (MPP), the unmixing algorithms attempt to recover $N$ pure spectral signatures (i.e., endmembers) and their corresponding spatial proportions (i.e., abundance maps). 
Mathematically, given a $M$-band RS images with $L$ pixels $\bX\in\mathbb{R}^{M\times L}$, the unmixing algotithms factorize it into the spectral signature matrix $\bE\in\mathbb{R}^{M\times N}$ and spatial abundance matrix $\bS\in\mathbb{R}^{N\times L}$ such that $\bX\approx\bE\bS$, where $N$ is typically assumed known.
Since the AVIRIS-level HSI is reconstructed from its Sentinel-2 counterpart over the same spatial location, they naturally refer to the consistent abundance maps. 
Accordingly, we compare the abundance maps (i.e., spatial proportions of pure materials) between a real Sentinel-2 MSI and its AVIRIS-level reconstruction for cross-validation.

The ROI (see Figure 7 of the main article) for this case study is located in Washington (WA), USA, and was captured in August 2019.
This ROI covers a spatial size of $256\times 256$, and primarily consists of river, farmland, and mountainous areas. 
We illustrate the true-color and false-color compositions of the real Sentinel-2 and the spatial prior image to evaluate their spectral relations, as demonstrated in Figure 7 of the main article.
The high-resolution and multiresolution compositions exhibit consistent color distributions, validating the effectiveness of the SRU provided by PriorNet.
In addition, we demonstrate that the high-resolution spatial prior images preserve a meaningful data structure, rather than being ``pretty pictures''.
Specifically, before performing the unmixing algorithm on the Sentinel-2 data and its AVIRIS-level reconstruction, the number of sources is estimated using the unsupervised model order selection method, namely minimum description
length (MDL) \cite{MDL}.
Developed by information theory, the MDL algorithm determines the optimal model order $N$ that results in the shortest code length for both real Sentinel-2 data and spatial prior image.
As demonstrated in Figure 7 of the main article, the code length with respect to the number of sources follows a consistent trend on both real Sentinel-2 data and spatial prior image, substantiating that the spatial prior image maintains a realistic multispectral structure. 

With the consistently shortest code length at $N:=4$, we employ a theoretically guaranteed unmixing algorithm [i.e., hyperplane-based Craig’s simplex
identification (HyperCSI) \cite{HyperCSI}] to recover four endmembers/abundance maps from both real Sentinel-2 data and its AVIRIS-level reconstruction, as demonstrated in Figure \ref{fig: unmixing_comparison}.  
In the results, we can tell that the most significant distinction lies in their spectral resolutions of recovered pure signatures (i.e., endmembers).
Due to the limited spectral resolution of the Sentinel-2 data (only 12 bands), it is challenging to identify materials using their discrete spectral signatures, even with the help of the unmixing algorithm.
In contrast, the AVIRIS-level HSI reconstructed by the proposed UALNet contains 186 high-quality and densely sampled bands across a broad spectrum range.
Such abundant spectral representation enables HyperCSI to recover the continuous hyperspectral endmembers, thereby facilitating more faithful RS applications
To further demonstrate the reliability and effectiveness of the proposed UALNet, we evaluate the spatial consistency of the recovered abundance maps between the Sentinel-2 data and its AVIRIS-level counterpart.
Because they correspond to the same geographic area, the reconstructed HSI in the high-spatial-resolution domain is expected to preserve material distributions aligned with the Sentinel-2 data in the coarse-spatial-resolution domain, rather than producing some spurious patterns artificially.
To verify this quantitatively, the abundance maps recovered from the AVIRIS-level reconstruction ($512\times 512$) are resized back to $256\times 256$ to match the spatial dimension with the real Sentinel-2 data, ensuring a quantitative evaluation.
Subsequently, we compute the cross-correlation coefficient for each pair of abundance maps, which is more robust and less sensitive (compared to distance-based metrics) to the scale distortions caused by resizing.
The averaged correlation across all abundance pairs (excluding the outlier minimum correlation) achieves 93.5388\%, validating the strong spatial consistency and the spatial reliability of the proposed UALNet.
In a nutshell, the proposed framework, namely unfolding adversarial learning network (UALNet), for the Sentinel-2 to AVIRIS-level reconstruction is experimentally validated to be applicable for real-world RS applications.
\section{Practical Applicability: Real-Data Analysis}\label{sec: realdata}
\begin{figure*}[t]
    \centering
    \centerline{\includegraphics[width=0.85\textwidth]{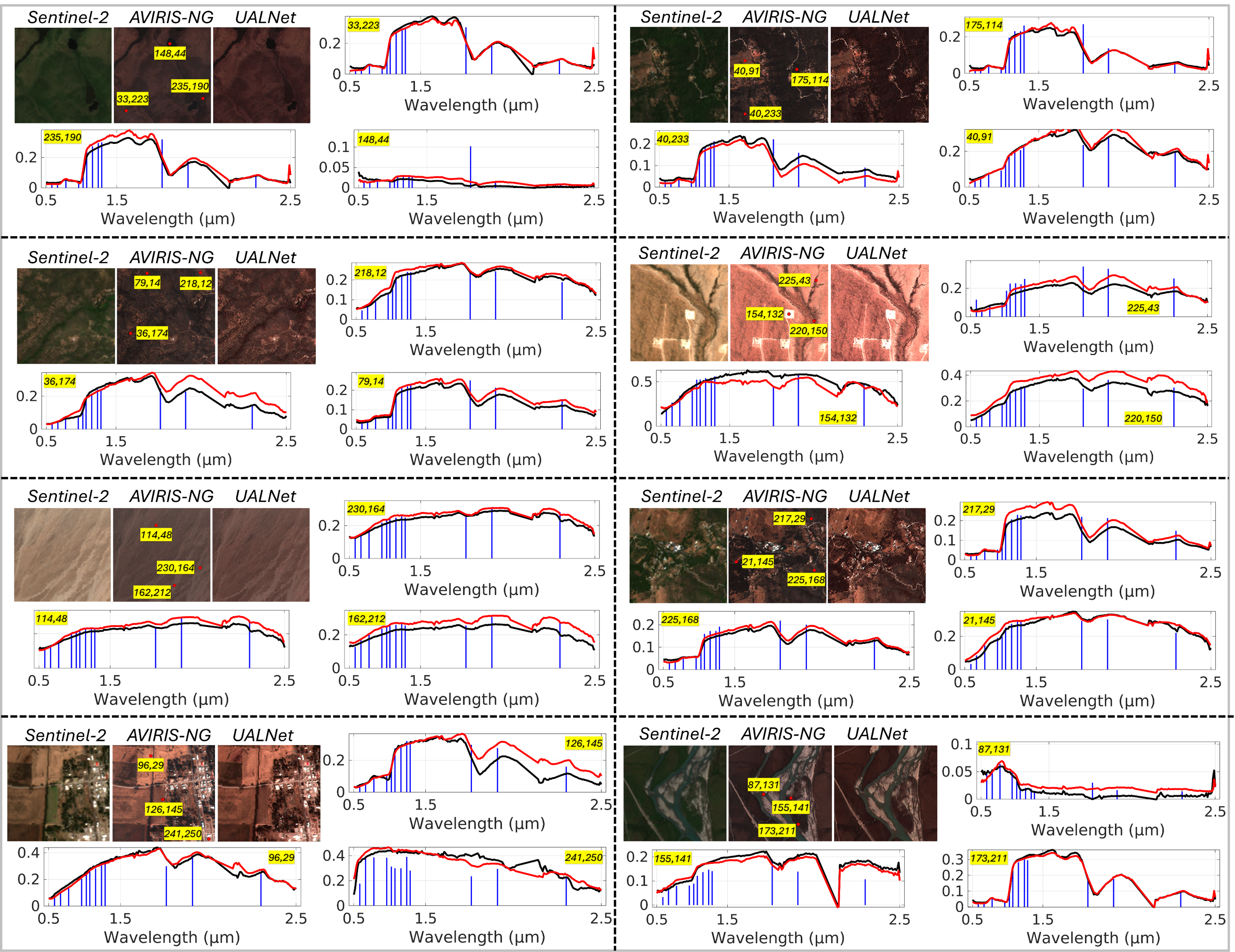}}
    \caption{Qualitative evaluations (bands 25 12 8 as RGB) on various real-world datasets, including Inuvik (Northwest Territories, Canada), Valley Springs (Calaveras, USA), Death Valley National Park (Inyo, USA), Pauls Valley (Garvin, USA), North Slope Borough (Alaska, USA), Carlsbad Caverns National Park (Eddy, USA), and Shingle Springs (El Dorado, USA). 
    The blue impulses denote Sentinel-2 signatures, while the black and red continuous curves are the real AVIRIS-NG and the UALNet-reconstructed signatures, respectively. 
    The UALNet trained on simulated data can yield promising reconstructions for real-world inputs.}\label{fig: real_data_signature_comparison}
\end{figure*}

Real-world data pairs are further collected to verify the simulation-to-real (S2R) generalization capability of the proposed UALNet.
Although the proposed UALNet is trained solely on simulated datasets, we apply it on real-world Sentinel-2 inputs (with $128\times 128$ pixels) \textbf{without additional fine-tuning}.
Then, we compare both false-color compositions and spectral signatures between the UALNet-reconstructed hyperspectral images and the corresponding real-world AVIRIS-NG dataset collected from diverse regions.
As shown in Figure \ref{fig: real_data_signature_comparison}, the reconstructed signatures (i.e., red curves) exhibit strong consistency across all significantly different real-world datasets.
The rigorous simulation training strategy enables robust S2R generalization.
These observations will be included in the main article during the revision stage.
